\documentclass[11pt,a4paper]{article}
\pdfoutput=1
\usepackage[utf8]{inputenc}
\usepackage{amsmath}
\usepackage{amsfonts}
\usepackage{array}
\usepackage{bbm}
\usepackage{amssymb}
\usepackage{cite}
\usepackage{graphicx}
\usepackage{enumitem}
\usepackage{appendix}
\usepackage{amsthm}
\usepackage{subcaption}
\usepackage[pdftex,pdfauthor={Ioannis Koukoutsidis},pdftitle={Age of Information in an Overtake-Free Network of Quasi-Reversible Queues}]{hyperref}
\author{Ioannis Koukoutsidis}
\title{Age of Information in an Overtake-Free Network of Quasi-Reversible Queues}

\theoremstyle{definition}

\newtheorem*{remark}{Remark}
\newcommand{\defeq}{\mathrel{\mathop:}=}
\begin{document}
\maketitle
\begin{abstract}
We show how to calculate the Age of Information in an overtake-free network of quasi-reversible queues, with exponential exogenous 
interarrivals of multiple classes of update packets and exponential service times at all nodes. Results are provided for any number of M/M/1 First-Come-First-Served (FCFS) queues in tandem, and for a network with two classes of update packets, entering through different queues in the network and exiting through the same queue. The main takeaway is that in a network with different classes of update packets, individual classes roughly preserve the ages they would achieve if they were alone in the network, except when shared queues become saturated, in which case the ages increase considerably.
\end{abstract}
\section{Introduction}
\label{Sect:Intro}
The Age of Information (AoI, or simply ``age'') equals the time from the generation of a piece of information, until that piece is received by its intended destination. In packet networks, we consider that information updates are received in packets, and at each time instant $t$, the destination observes an age $H(t)=t-u(t)$, where $u(t)$ is the generation time of the last packet received.\footnote{We use the letter $H$ for the AoI, from the greek word `$H\hspace{-0.2em}\lambda\iota\kappa\iota^{\hspace{-0.22em}\prime}\hspace{-0.2em}\alpha$', which means `Age'.} If the process $H(t)$ is ergodic, then its expectation $E[H(t)]$ converges to the average $\bar{H}$ of the ages of all packets, as seen by the destination when the packets are received. The latter metric will be meant when referring to the AoI metric in this paper. Packets may also be of different types or classes, in which case we are interested in the age $\bar{H}^c$ separately for each class $c$.

The AoI can serve as a metric in numerous applications, where we are interested in the freshness of the received information. For example, in the Internet of Things, where sensor devices can transmit updates of environmental parameters, or the values of technical parameters such as location and velocity in autonomous vehicles; in the storage of data in computer systems, where we are interested in the freshness of imformation in the cache memory, or in robotics and control systems, where the fast feedback of information plays a prominent role.

These applications justify the necessity for the AoI as a metric, but perhaps its importance stems more from the fact that it motivates a redesign of computer and networking systems, to also take it into account, as the traditional metrics of throughput and delay were shown from the very start to poorly reflect freshness \cite{kaul2012real}. Moreover, information freshness trade-offs with energy efficiency, this also motivating the scheduling of transmissions of energy-constrained devices to take AoI into account \cite{wu2017optimal}. An extensive survey of conducted research and the numerous applications of the AoI metric was recently published in \cite{yates2020age}.

Mathematically, the main tool for theoretically calculating the AoI is queueing theory. Update packets are seen as arriving to a queue, where they wait for processing by a monitor, which reads the related information. So far, results have mainly been derived for single-server queues, with the exception of \cite{yates2018age}, where the AoI was calculated for M/M/1 queues in tandem under Last-Come-First-Served (LCFS) service discipline with service preemption, and the more general method in \cite{yates2018networks} (see Sect.~\ref{sect:rel_work} for a description of related work). The method in \cite{yates2018networks} can theoretically be applied for any network in which the movements of updates are described by a finite-state continuous-time Markov chain, but it is a matrix-based method which can lead to combinatorial explosion for systems with a large number of states, such as networks with infinite queues.

In this paper, we derive results for the AoI in networks, starting from the initial analysis in \cite{kaul2012real}, but presenting it in a new, more simplified manner, and then relying on classical queueing theory results from \cite{melamed1982sojourn} and \cite{walrand1988introduction}, which were derived more than three decades ago.
The model (in Sect.~\ref{sect:model}) and analysis (in Sect.~\ref{Sect:o-f path}) allow to calculate the AoI for different types or classes of update packets, both at the output of each node, where the packets pass from, and at the exit of the network. 

The concept that allows us to extend the calculation of the AoI to a queueing network is that of a \textit{quasi-reversible} queue.
A Markovian queue\footnote{A queue is called Markovian if its state process is a stationary and ergodic Markov chain.} is called quasi-reversible if the state of the process at time $t$ is independent of the arrival process after time $t$ and is also independent
of the departure process prior to time $t$ \cite{nelson1993mathematics}. The important property of a quasi-reversible queue, which 
will be of great use here, is that Poisson streams passing through the queue are statistically unchanged, hence producing Poisson output streams with the same rate. This is called the \textit{Poisson-in-Poisson-out property}. 

When quasi-reversible queues are connected into a network, the network is called a \textit{quasi-reversible network}. The network itself possesses the Poisson-in-Poisson-out property; however, whether or not internal traffic is Poisson depends on the topology of the network \cite{walrand1988introduction}. More specifically, internal traffic is Poisson in \textit{overtake-free}, or \textit{order-preserving networks}. These are networks where the order of transmission is preserved and jobs sent later in time cannot arrive at the destination earlier than previously sent jobs. Note that if there is a possibility for update packets to arrive through multiple routes to a destination, then a packet that was sent earlier may arrive later than a subsequent packet, which carries fresher information. The destination could simply discard the packet that arrived later, but this significantly complicates the analysis. Thus, we confine ourselves to overtake-free paths in this paper. 

We extend the AoI results in the literature by calculating the AoI for \textit{any} number of M/M/1 queues in tandem, as well as a simple network with two classes of update packets, where the different class update packets do not \textit{directly} share an output queue, but pass from other nodes first (see Sect.~\ref{Sect:Examples}). The use of FCFS M/M/1 queues might appear simplistic, but apart from its theoretical value, it also has practical significance in networks of queues with multiple classes of update packets, where different class packets should share common resources, without one suffocating the other. More specifically, it was shown in \cite{yates2018SHS} that for two classes of packets directly sharing an output M/M/1 queue, the FCFS discipline can minimize the sum AoI of the two classes for a large range of load values, outperforming LCFS disciplines both with and without preemption in service.\footnote{The analysis in \cite{yates2018SHS} contained an error, which originated from \cite{yates2012real} and was only recently corrected in \cite{kaul2020timely}. However, as shown in \cite{kaul2020timely}, the error is numerically small and the qualitative conclusions for FCFS and LCFS queues remain valid.}  
\section{Related work}
\label{sect:rel_work}
The basic method for calculating the AoI in a single server queue with First-Come-First-Served (FCFS) service discipline was shown in \cite{kaul2012real}, where the authors derived results for M/M/1, M/D/1 and D/M/1 queues. This paper (similarly to the majority of other works so far) builds upon this basic method. Another fundamental paper was \cite{costa2014age}, where the authors derived results for M/M/1/1 (at most one packet being served, no packets waiting) and M/M/1/2$^*$ (a variation of M/M/1/2, with at most one packet being served and one packet waiting, and the additional assumption that the packet waiting is discarded if a new packet arrives while the server is busy) queues with FCFS service discipline. This paper also included the definition of the \textit{peak AoI}, or \textit{peak age} metric, which reflects the average maximum age seen by the destination, prior to receiving updates.

The intuition that waiting may not be efficient when we are interested in receiving update packets as soon as possible, led to the exploration of LCFS service disciplines. Results for a single server M/M/1 queue were derived in \cite{kaul2012status}, where the authors showed that LCFS, with or without service preemption, always outperforms FCFS for a single class of update packets. Enforcing packet deadlines, or timeouts, after which an update packet waiting in the queue is discarded, was studied in \cite{kam2018age} for an M/M/1/2 queue, showing that using a deadline can outperform both the M/M/1/1 and M/M/1/2 without deadline.

Results for multiple sources sharing a single-server M/M/1 queue with FCFS service discipline were first derived in \cite{yates2012real}. This also produced insights into how a source can choose its update rate in the presence of interfering traffic, basically showing that the minimum AoI is achieved at a smaller update rate, compared to when only one source sends packets to the queue. For the peak AoI, results for more general multi-class M/G/1 and M/G/1/1 queues were presented in \cite{huang2015optimizing}. 

A breakthough in the method of calculating the AoI was achieved in \cite{yates2018age}, where the author used a Stochastic Hybrid System (SHS) to model the system and derived AoI results for M/M/1 queues in tandem, under LCFS service discipline with service preemption. The use of SHS leads to a general method for calculating the AoI for a variety of queues with different service disciplines. In \cite{yates2018SHS}, the authors showed the application of the method for calculating the AoI for two-sources sharing an M/M/1 queue under FCFS and LCFS service disciplines. The technique was presented more generally in \cite{yates2018networks} for any network described by a finite-state continuous-time Markov chain, and led to a system of ordinary linear differential equations that describe the temporal evolution of the moments and moment-generating functions of the age process. For a line network of preemptive memoryless servers and Poisson input, \cite{yates2018networks} showed that the age at a node is identical in distribution to the sum of the interarrival interval and the service times at the preceding nodes (all exponentially distributed). 

\section{The model}
\label{sect:model}
We consider a quasi-reversible network consisting of a set of nodes $M=\{1,\dots,m\}$ and a fictitious node 0 that represents both the source and the sink of the network. The network admits different types, or classes of update packets; these will be the jobs circulating in the network, and may correspond to different applications that share the network for transmission. External update packets of type $c$ 
arrive at the network from the source node 0 as a Poisson process with rate $\lambda_0^{c}$. Each arrival joins node $i$ with
probability $r_{0,i}^c$. On service completion at node $i$, it is routed to node $j$ with probability $r_{i,j}^c$, or leaves the network with probability $r_{i,0}^c=1-\sum_{j\neq 0}r_{i,j}^c$. Generally, an update packet of type $c$ follows a path or itinerary $r^c\defeq(r_1^c,\dots,r_n^c)$, defined by the sequence of nodes $r_1^c,\dots,r_n^c$ visited by the
packet, from its entry in the network through node $r_1^c$, until its exit from the network via node $r_n^c$. The number of nodes $n$ is called the \textit{length} of the path. When the path and the packet type are obvious, we will simply denote a path by its start and end nodes, as $r_1\to r_n$.

In the queueing network, the total equilibrium rate of type-$c$ update packets through node $i$ (including both external arrivals and internal transitions) can be found by solving the system of \textit{traffic equations}:
\begin{equation}
	\lambda_i^c=\lambda_0^c r_{0,i}^c+\sum_{j\neq i}\lambda_j^c r_{j,i}^c, \quad i,j=1,\dots,m\;.
\end{equation}
The rate at which update packets of type $c$ exit the network from node $i$ ($i=1,\dots,m$) is $\sum_i \lambda_i^c r_{i,0}^c$. Additionally, the total rate of update packets (of all types) in node $i$ is $\lambda_i=\sum_c \lambda_i^c$.

We assume that all nodes have a FCFS queueing discipline and do not admit batch processing, neither go on vacations. The service times of all packets at each node $i$ are exponentially distributed with parameter $\mu_i$. Additionally, the network is assumed to contain only overtake-free paths.
An overtake-free path is defined as a path:
\begin{itemize}
	\item where every node is overtake-free (this is already assumed by the FCFS service discipline),
	\item which is cycle-free, i.e. every node in the path is distinct, and
	\item which does not contain any forward short-circuits, i.e. packets sharing parts of an itinerary in the
	forward direction cannot overtake each other by taking a shorter route (a route which intersects two non-contiguous
	nodes of another packet's route).
\end{itemize}
For a more rigorous definition of overtake-free paths, see \cite{melamed1982sojourn}.

It is straightforward to understand the need for overtake-free paths for simplifying the calculation of the AoI:
as long as the path is overtake-free, the age of an update packet, as calculated in some point in the forward path, 
does not depend on update packets behind it. Otherwise, other packets that overtake it will affect its age.
Moreover, similarly to what was noted in \cite{melamed1982sojourn} for the calculation of sojourn times, backward short circuits 
are immaterial for the calculation of the AoI,
since packets along such paths do not change the order in the forward direction.
\section{AoI in an overtake-free path}
\label{Sect:o-f path}
\begin{figure}[!htb]
	\centering
	\includegraphics[scale=.7]{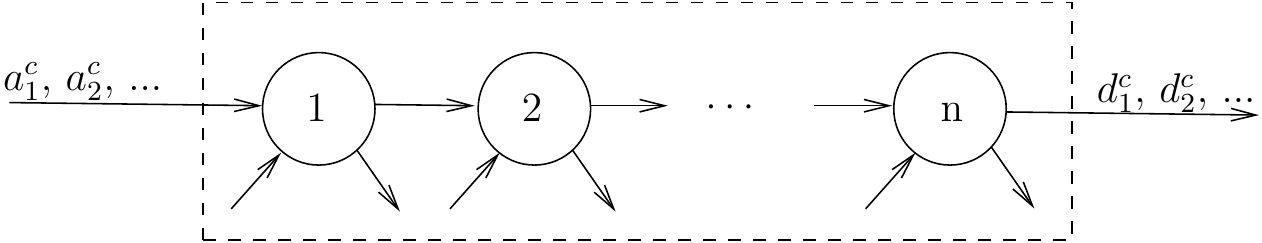}
	\caption{A view of a path as a ``black-box''\label{pathview}}
\end{figure}

Without loss of generality, we will calculate the AoI for a type of update packets at the output of an overtake-free path, by looking at the path as a black box and only being interested in the input and output processes. Consider the path of length $n$ shown in Fig.\ref{pathview}, where update packets of type $c$ arrive at the first node in the path as a Poisson process at time instants $(a_1^c,a_2^c,\dots)$, and exit the path at time instants $(d_1^c,d_2^c,\dots)$. There may also be traffic of other types sharing the path, either wholly or partially, depicted by the diagonal arrows in the figure. The total traffic at each node $j$ ($j=1,\dots,n$) is $\lambda_j$, and the service rate is $\mu_j$. Given that $\lambda_j<\mu_j$ (so that all queues are stable), we will calculate the AoI at the output of the path. We will further assume that all update packets arrive fresh at the entry of the path, i.e. their initial ages are zero.

For an overtake-free path, it was shown in \cite{melamed1982sojourn} that the end-to-end sojourn time of a customer is distributed in the limit as a sum of independent exponential sojourns at each node $j$, ($j=1,\dots,n$), with respective parameters $\mu_j-\lambda_j$. That is, the sojourn time of a job in the network depends on its type only through the itinerary followed, and sojourn time of different type jobs in the same node have the same distribution.

\begin{figure}[!tb]
	\centering
	\includegraphics[scale=.7]{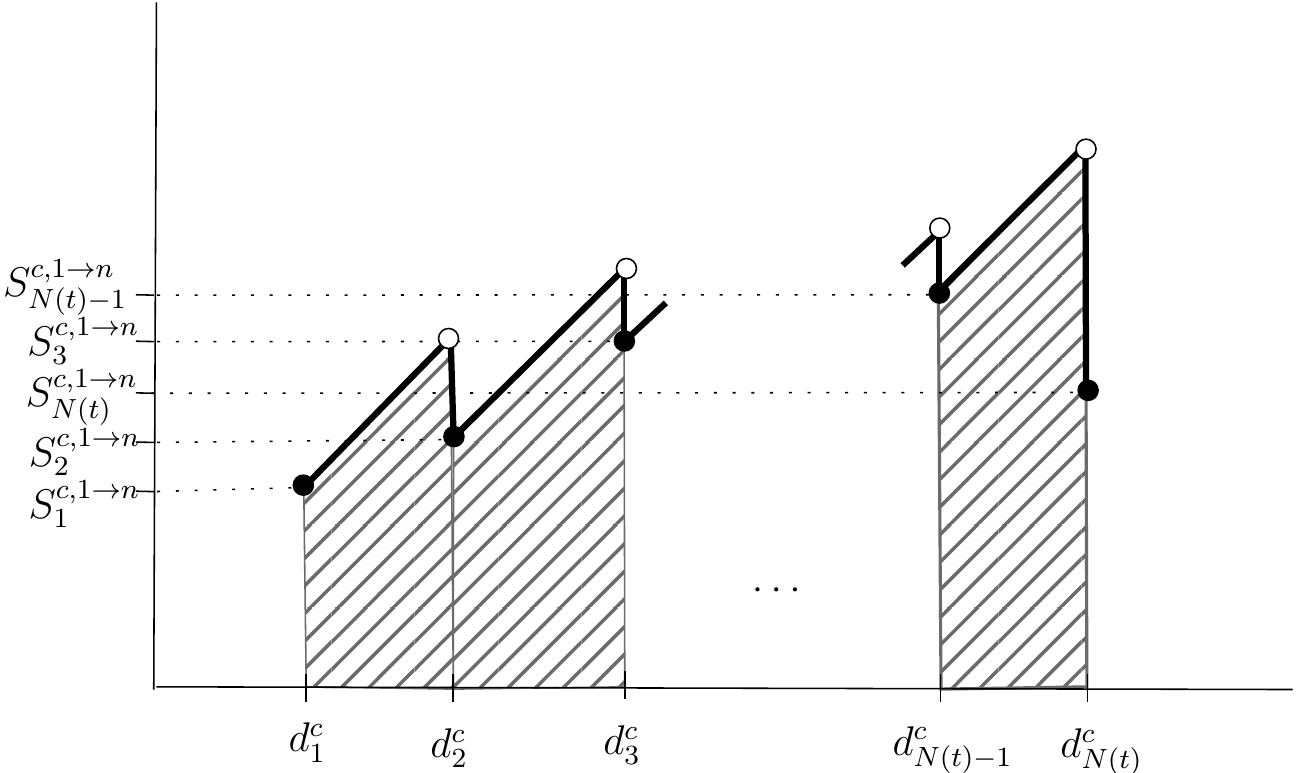}
	\caption{Sample run of the age update process\label{sawtooth}}
\end{figure}
To illustrate the method of calculation, consider an example run of the age process at the output of the path, as shown in Fig.~\ref{sawtooth}. At time $d_i^c$, the age of the type-$c$ update packet received at the output of the path is equal to its end-to-end sojourn time $S_i^{c,1\to n}$ through the whole path.

The AoI at the receiver is equal to the time average
\begin{equation}
	\bar{H}^c=\lim_{t\to\infty}\frac{1}{t}\left(\sum_{i=1}^{N(t)-1}(S_i^{c,1\to n}+(d_{i+1}^c-d_i^c)/2)(d_{i+1}^c-d_i^c)\right)\;,
\end{equation}
where $N(t)$ is the number of packet departures in time $t$ (the receiver perceives a mean age $S_i^{c,1\to n}+(d_{i+1}^c-d_i^c)/2$ in the interval $[d_i^c,d_{i+1}^c]$).

Similarly, we can define the left and right limits bounding the AoI:
\begin{subequations}
	\begin{equation}
		\bar{H}_{left}^c=\lim_{t\to\infty}\frac{1}{t}\left(\sum_{i=1}^{N(t)-1}S_i^{c,1\to n}(d_{i+1}^c-d_i^c)\right)\;,
		\label{left_limit}
	\end{equation}
	\begin{equation}
		\bar{H}_{right}^c=\lim_{t\to\infty}\frac{1}{t}\left(\sum_{i=1}^{N(t)-1}(S_i^{c,1\to n}+d_{i+1}^c-d_i^c)(d_{i+1}^c-d_i^c)\right)\;.
		\label{right_limit}
	\end{equation}
\end{subequations}
In Fig.~\ref{sawtooth} the age process is depicted as the sawtooth curve in bold,
and is right-continuous with left limits. The right limits (solid black circles) average to the lower (left) bounds of the age process (\ref{left_limit}), while the left limits (solid white circles) average to the upper (right) bounds of the update process (\ref{right_limit}). 

It is straightforward to see that
\begin{subequations}
	\begin{equation}
		\bar{H}^c=\bar{H}_{left}^c+\lim_{t\to\infty}\frac{1}{t}\left(\sum_{i=1}^{N(t)-1}(d_{i+1}^c-d_i^c)^2/2\right)\;,
	\end{equation}
	\begin{equation}
		\bar{H}_{right}^c=\bar{H}_{left}^c+\lim_{t\to\infty}\frac{1}{t}\left(\sum_{i=1}^{N(t)-1}(d_{i+1}^c-d_i^c)^2\right)\;.
	\end{equation}
	\label{left_right_ages_analytical}
\end{subequations}
The $D_i^c\defeq d_{i+1}^c-d_i^c$ correspond to the interdeparture times of the type-$c$ update packets exiting the path, and, for an overtake-free path of quasi-reversible queues, in the limit $t\to\infty$ successive $D_i^c$ $(i=1,\dots,N(t)-1)$ are independent and have the same exponential distribution. Therefore it holds with probability 1 that $\lim_{t\to\infty}N(t)/t=1/E[D_i^c]$. Moreover, since the interdeparture process is ergodic, 
\begin{align*}
	&\lim_{t\to\infty}\frac{1}{t}\left(\sum_{i=1}^{N(t)-1}(d_{i+1}^c-d_i^c)^2\right)\\
	=&\lim_{t\to\infty}\frac{N(t)-1}{t}\left(\frac{\sum_{i=1}^{N(t)-1}(d_{i+1}^c-d_i^c)^2}{N(t)-1}\right)\\
	=& E[(D_i^c)^2]/E[D_i^c]\;.
\end{align*} 
Therefore, equations (\ref{left_right_ages_analytical}) become:
\begin{subequations}
	\begin{equation}
		\bar{H}^c=\bar{H}_{left}^c+\frac{E[(D_i^c)^2]}{2E[D_i^c]}\;,\label{mean_AoI}
	\end{equation}
	\begin{equation}
		\bar{H}_{right}^c=\bar{H}_{left}^c+\frac{E[(D_i^c)^2]}{E[D_i^c]}\;.\label{A_right}
	\end{equation}
	\label{left_right_ages_lambda}
\end{subequations}
Similarly, from (\ref{left_limit}):
\begin{equation}
	\bar{H}_{left}^c=\frac{E[S_i^{c,1\to n} D_i^c]}{E[D_i^c]}\;.
	\label{A_left}
\end{equation}
In summary, the AoI of a type of update packets only depends on the departure rate of that type and the expected value of the product between the sojourn time of a packet and the interdeparture interval between the departure of that type of packet and the one of the same type following it in the path.

\begin{remark}
	Note that $\bar{H}_{right}^c$ is \textit{not} equal to the \textit{peak age}, as defined in \cite{costa2014age}. For an M/M/1 FCFS queue, the peak age equals $E[A^c]+E[S^c]=E[D^c]+E[S^c]$, where $A^c$, $D^c$ are the interarrival and interdeparture intervals of class-$c$ packets, respectively \cite{huang2015optimizing}. From the analysis that follows, one can readily derive that $\bar{H}_{right}^c$ is a looser upper bound to $\bar{H}^c$ than the peak age, becoming closer to the peak age for heavy traffic.
\end{remark}

We calculate the $E[S_i^{c,1\to n} D_i^c]$ by noting that, since the network (consisting of all queues in the path) is quasi-reversible, the distribution of interarrivals is the same as the distribution of interdepartures.\footnote{The system is also time-reversible, since the interval between the departure of a packet and the one that followed it in the forward-time system corresponds to the interval between the arrival of a packet and the one that preceded it in the reverse-time system.} Denoting the interarrival interval $(a_i^c-a_{i-1}^c)$ by $A_i^c$, it therefore holds that $D_i^c\sim A_i^c$, and $S_i^{c,1\to n} D_i^c\sim S_i^{c,1\to n} A_i^c$.

When the parameter of the exponential interarrival distribution for type-$c$ update packets is $\lambda_c$, we have $E[(D_i^c)^2]=2/\lambda^2_c$ and (\ref{mean_AoI}), (\ref{A_right}), (\ref{A_left}) become:
\begin{subequations}
	\begin{equation}
		\bar{H}^c=\bar{H}_{left}^c+\frac{1}{\lambda_c}\;,\label{A_mean_lambda}
	\end{equation}
	\begin{equation}
		\bar{H}_{right}^c=\bar{H}_{left}^c+\frac{2}{\lambda_c}\;,\label{A_right_lambda}
	\end{equation}
	\begin{equation}
		\bar{H}_{left}^c=\lambda_c E[S_i^{c,1\to n} D_i^c]\;.
		\label{A_left_lambda}
	\end{equation}
\end{subequations}

Denoting by $S_i^{c,j}$ the sojourn time of type-$c$ update packet $i$ at node $j$, we have $S_i^{c,1\to n}=\sum_{j=1}^n S_i^{c,j}$. We decompose $S_i^{c,j}$ to $W_i^{c,j}$ and $X_i^{c,j}$, the waiting time and service time of this packet at node $j$. When dealing with distributions or expectations of $S_i^{c,j}$, $W_i^{c,j}$ and $X_i^{c,j}$, we can drop the index $c$, because these distributions are independent of packet type. Since each of the successive sojourn, interarrival and service times are i.i.d, their expectations are the same for all successive packets, and the subscripts $i$ can also be dropped for notational simplicity.
We have:
\begin{align}
	E[S^{c,1\to n} D^c]&=E[S^{c,1\to n} A^c]=\sum_{j=1}^n E[S^{j} A^c]\nonumber\\
	&=\sum_{j=1}^n E[(W^{j}+X^{j}) A^c]\nonumber\\
	&=\sum_{j=1}^n (E[W^{j} A^c]+E[X^{j}]E[A^c])\;.
	\label{Si..nDi}
\end{align}
(The last step occurs because of the independence of service times and interarrival times.) 

Consider the node $j$ with input (and output) rate of type-$c$ update packets $\lambda_{c,j}$, total arrival rate (of all-type update packets) $\lambda_j$ and service rate $\mu_j$ (irrespective of the type of update packets). Denote the total load at queue $j$ by $\rho_j=\lambda_j/\mu_j$, and the load of class-c by $\rho_{c,j}=\lambda_{c,j}/\mu_j$. Obviously $\sum_c\rho_{c,j}=\rho_j$.

Applying Lemma 1 in  \cite{yates2012real}, the $E[W^{j} A^c]$ is given by:\footnote{This also corrects a mistake in the calculation of $E[W^{j}A^c]$ in \cite{koukou20}, which didn't fully take into account the presence of other class packets, resulting in lower age values.}
\begin{align}
	E[W^{j}A^c]&=\frac{1}{\mu_j^2}\left[\frac{\rho_{c,j}(1-\rho_j(\rho_j-\rho_{c,j}))}{(1-\rho_j)(1-(\rho_j-\rho_{c,j}))^3}+\frac{\rho_j-\rho_{c,j}}{\rho_{c,j}(1-(\rho_j-\rho_{c,j}))} \right]\;.\label{wiai}
\end{align}

Combining (\ref{Si..nDi}) and (\ref{wiai}), we get
\begin{gather}
	E[S^{c,1\to n} D^c]=\sum_{j=1}^n \bigg[\frac{1}{\mu_j^2}\left[\frac{\rho_{c,j}(1-\rho_j(\rho_j-\rho_{c,j}))}{(1-\rho_j)(1-(\rho_j-\rho_{c,j}))^3}+\frac{\rho_j-\rho_{c,j}}{\rho_{c,j}(1-(\rho_j-\rho_{c,j}))} \right]\nonumber\\
	+\frac{1}{\mu_j\lambda_c}\bigg]\;.
\end{gather}

Finally, from (\ref{A_left_lambda}) and (\ref{A_mean_lambda}), we get the AoI for type-$c$ update packets at the output of the path:
\begin{gather}
	\bar{H}^c=\sum_{j=1}^n \left[\frac{\lambda_c}{\mu_j^2}\left[\frac{\rho_{c,j}(1-\rho_j(\rho_j-\rho_{c,j}))}{(1-\rho_j)(1-(\rho_j-\rho_{c,j}))^3}+\frac{\rho_j-\rho_{c,j}}{\rho_{c,j}(1-(\rho_j-\rho_{c,j}))} \right]\right]\nonumber\\
	+\sum_{j=1}^n \frac{1}{\mu_j}+\frac{1}{\lambda_c}\;.
	\label{H_path}
\end{gather}

In the next section we provide some illustrative examples.  
\section{Examples}
\label{Sect:Examples}
\subsection{Queues in tandem}
Consider the simple network shown in Fig.~\ref{pathview}, where $n$ queues are connected in tandem and there is only one type of update packets, which arrive at the network with rate $\lambda$. For simplicity, it is assumed that the queues have the same service rate $\mu$. In order for the queues to be stable, we must have $\rho=\lambda/\mu<1$. It is noted that so far results have only been taken for a network of M/M/1 preemptive LCFS queues in tandem \cite{yates2018age}.

Taking $\lambda_c=\lambda_j=\lambda$ and $\mu_j=\mu$ $\forall j$, (\ref{H_path}) becomes:
\begin{equation}
	\bar{H}=\frac{n\rho^2}{\mu-\lambda}+\frac{n}{\mu}+\frac{1}{\lambda}\;.
\end{equation}
\begin{figure}[!b]
	\centering
	\includegraphics[scale=1]{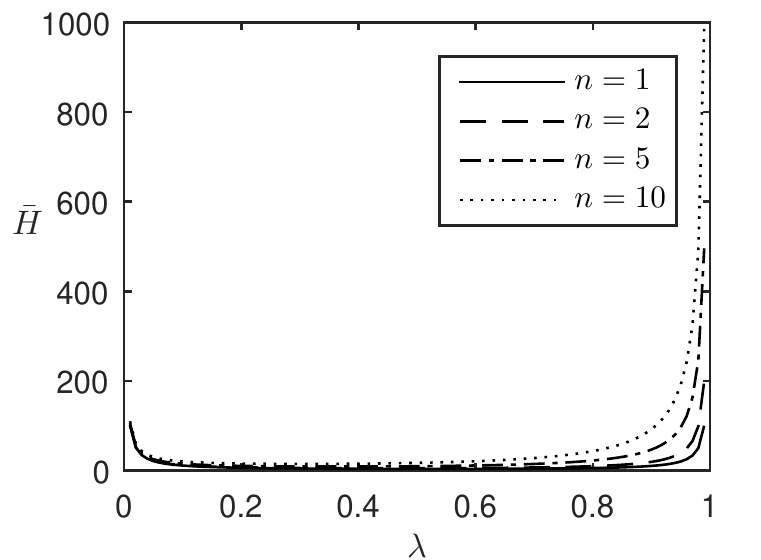}
	\caption{AoI for $n$ identical M/M/1 queues in tandem, for $\mu=1$ and varying values of the arrival rate $\lambda$.\label{plot_tandem}}
\end{figure}

\figurename~\ref{plot_tandem} shows results for the mean AoI for $\mu=1$ and varying values of $\lambda$, $n$. 
The same pattern appears, as in the case of a single queue: the minimum is achieved for some intermediate load value.
This value also decreases with the number of nodes; for $n=1$, the minimum is achieved approximately at $\rho=0.53$, for $n=2$ at $\rho=0.46$, for $n=5$ at $\rho=0.37$ and for $n=10$ at $\rho=0.31$. We also remark that the minimum AoI does not increase significantly with the number of nodes. The effect of the added queues is more apparent as the load increases, and the age can increase tremendously for large values. In the figure, the AoI increase for $\rho=0.99$ is equal to 891 time units between $n=1$ and $n=10$.
\subsection{A simple network with two classes}
\label{subsect:example_two_classes}
\begin{figure}[!htb]
	\centering
	\includegraphics[scale=.7]{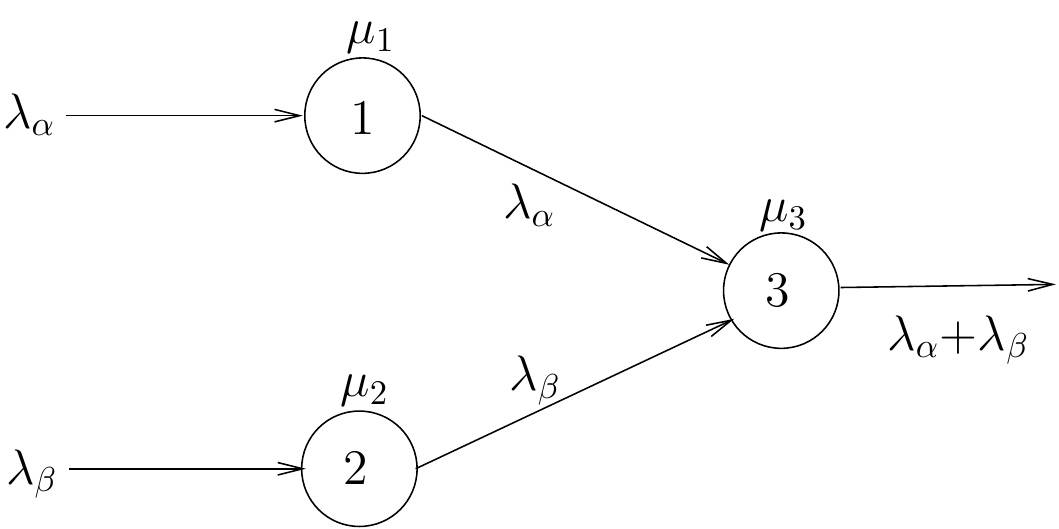}
	\caption{A simple network with two classes\label{fig:two_classes}}
\end{figure}

Consider the network with two classes shown in Fig.~\ref{fig:two_classes}. There are two types of update packets $\alpha$ and $\beta$, entering the network as Poisson processes with rates $\lambda_{\alpha}$ and $\lambda_{\beta}$, through nodes 1 and 2 respectively. Both types of packets exit the network through node 3. The service times at each node $i$ are exponential with mean $\mu_i^{-1}$. We will calculate the AoI for each class of update packets at the exit of the network. It is noted that the case where two classes of traffic directly share an output queue (without passing through a network) was studied in \cite{yates2012real}.

We assume that the stability conditions $\lambda_{\alpha}<\mu_1$, $\lambda_{\beta}<\mu_2$ and $\lambda_{\alpha}+\lambda_{\beta}<\mu_3$ hold. Since all queues are quasi-reversible and all paths are overtake-free, internal traffic is Poisson and the rates of class-$\alpha$ and class-$\beta$ update packets at the output of nodes 1 and 2 are also $\lambda_{\alpha}$ and $\lambda_{\beta}$, respectively, and the output rate at node 3 is $\lambda_{\alpha}+\lambda_{\beta}$.

From (\ref{H_path}), substituting $c\equiv\alpha$, we can calculate the AoI of class-$\alpha$ update packets at the output of node 3:
\begin{equation}
	\bar{H}^{\alpha}=\frac{\rho_1^2}{\mu_1-\lambda_{\alpha}}+\frac{\lambda_{\alpha}}{\mu_3^2}\left(\frac{\rho_{\alpha,3}(1-\rho_3\rho_{\beta,3})}{(1-\rho_3)(1-\rho_{\beta,3})^3}+\frac{\rho_{\beta,3}}{\rho_{\alpha,3}(1-\rho_{\beta,3})} \right)+\frac{1}{\mu_1}+\frac{1}{\mu_3}+\frac{1}{\lambda_{\alpha}}\;,
\end{equation}
where $\rho_1\equiv\rho_{\alpha,1}=\lambda_{\alpha}/\mu_1$ is the load at queue 1, $\rho_3=(\lambda_{\alpha}+\lambda_{\beta})/\mu_3$ the load at queue 3, and $\rho_{\alpha,3}=\lambda_{\alpha}/\mu_3$, $\rho_{\beta,3}=\lambda_{\beta}/\mu_3$ are the loads at queue 3 of classes $\alpha$, $\beta$, respectively. 

By symmetry, the AoI of class-$\beta$ update packets in the path $2\to3$, $\bar{H}^{\beta}$, can then be found by interchanging $\alpha$ and $\beta$, and $\mu_1$ and $\mu_2$ in the expression for $\bar{H}^{\alpha}$, and also defining $\rho_2\equiv\rho_{\beta,2}=\lambda_{\beta}/\mu_2$:
\begin{equation}
	\bar{H}^{\beta}=\frac{\rho_2^2}{\mu_2-\lambda_{\beta}}+\frac{\lambda_{\beta}}{\mu_3^2}\left(\frac{\rho_{\beta,3}(1-\rho_3\rho_{\alpha,3})}{(1-\rho_3)(1-\rho_{\alpha,3})^3}+\frac{\rho_{\alpha,3}}{\rho_{\beta,3}(1-\rho_{\alpha,3})} \right)+\frac{1}{\mu_2}+\frac{1}{\mu_3}+\frac{1}{\lambda_{\beta}}\;.
\end{equation}

\begin{figure}[!htb]
	\centering
	\begin{subfigure}[t]{0.5\textwidth}
		\includegraphics[width=\linewidth]{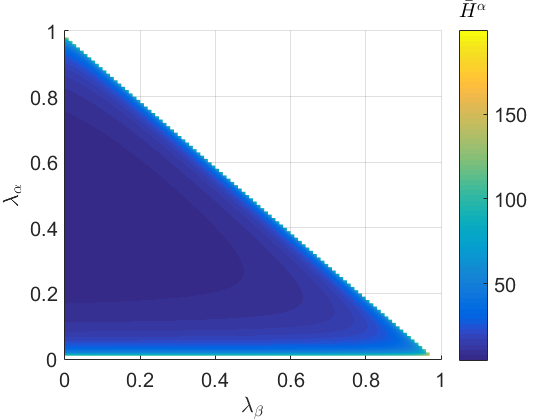}
		\caption{AoI of class-$\alpha$ packets\label{fig:two_classes_results}}
	\end{subfigure}
	\begin{subfigure}[t]{0.49\textwidth}
		\centering
		\includegraphics[width=\linewidth]{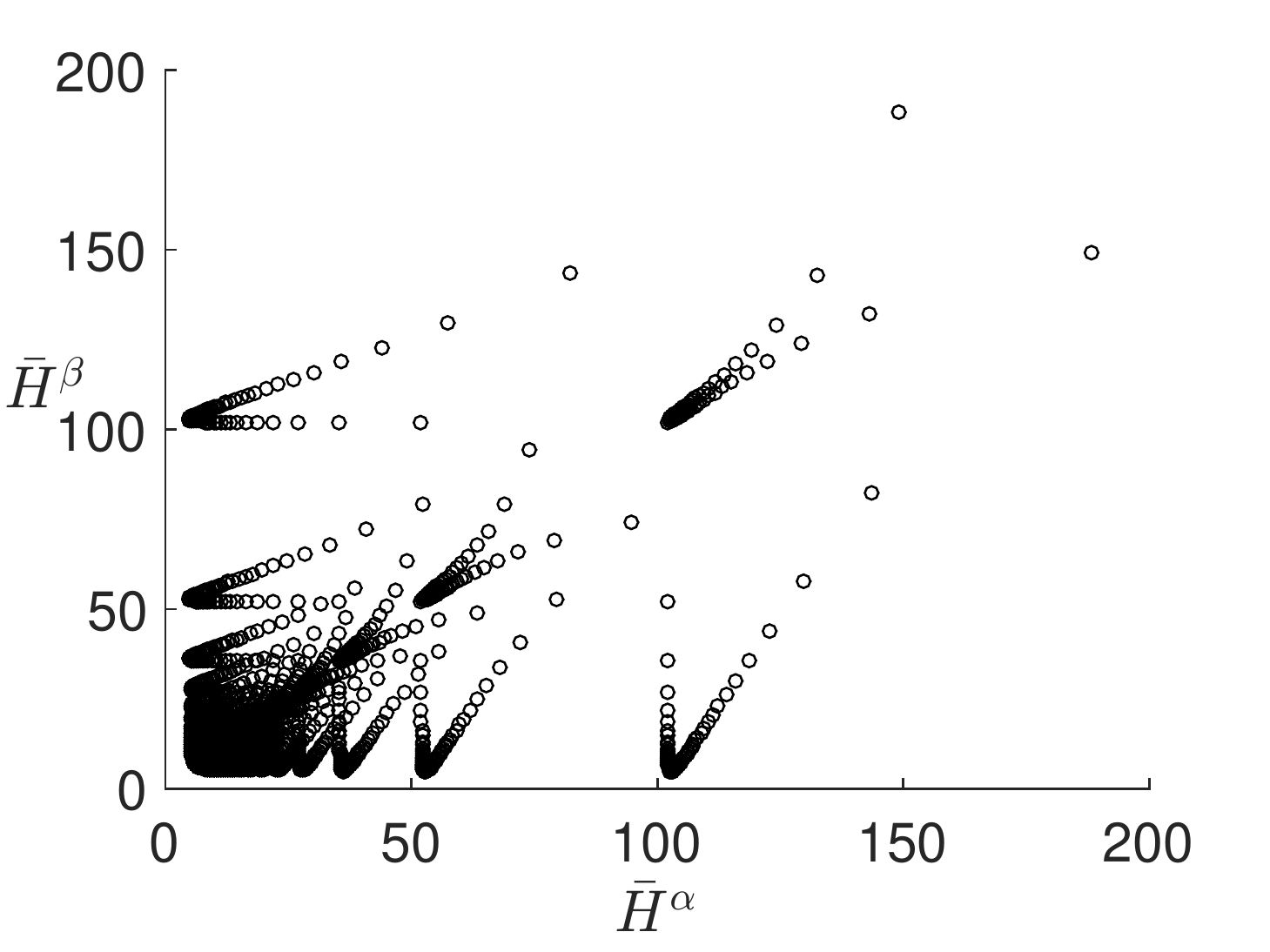}
		\caption{Scatter plot of $\bar{H}^{\alpha}$, $\bar{H}^{\beta}$\label{fig:two_classes_H1H2}}
	\end{subfigure}
	\caption{Results for the AoI in the network of Fig.~\ref{fig:two_classes}, for $\mu_i=1$ and varying $\lambda_{\alpha}$, $\lambda_{\beta}$.}
\end{figure}

Fig.~\ref{fig:two_classes_results} shows how the AoI of update packets in the output of the path $1\to 3$ varies in the region of $\lambda_{\alpha}$, $\lambda_{\beta}$ values (the service rates $\mu_i$ are fixed at 1 for all queues). The AoI is represented by color intensities in the different parts of the feasible region.
It can be seen that $\bar{H}^{\alpha}$ primarily depends on the value of $\lambda_{\alpha}$; the arrival of class-$\beta$ packets effects by increasing $\bar{H}^{\alpha}$, but only when $\lambda_{\beta}$  is high enough, so that the queue is close to saturation.

Fig.~\ref{fig:two_classes_H1H2} shows a scatter plot of $\bar{H}^{\alpha}$, $\bar{H}^{\beta}$ values in the region defined by $\lambda_{\alpha}$, $\lambda_{\beta}$. We notice that there are subregions where the AoI of one class of update packets stays relatively stable while the other's increases, and subregions where they both increase. This is in agreement with Fig.~\ref{fig:two_classes_results}, where increasing the arrival rate of one class always effects on the AoI of that class, but the other class is affected only when the output queue is close to saturation.

The minimum of $\bar{H}^{\alpha}$ equals 4.96 and is achieved when $\rho_1=0.46$ and $\rho_2=0$ (and vice-versa for $\bar{H}^{\beta}$). This is anticipated, as it coincides with the minimum $\rho$ value for $n=2$ in the case of queues in tandem. The minimum value for $\bar{H}^{\alpha}+\bar{H}^{\beta}$ equals 12.86 and is achieved for $\rho_1=\rho_2=0.3$, corresponding to $\bar{H}^{\alpha}=\bar{H}^{\beta}=6.43$. This is higher than the case where classes $\alpha$ and $\beta$ directly share the output queue, which was studied in \cite{yates2012real}, \cite{kaul2020timely}, and where the sum was minimized again at $\rho_1=\rho_2\approx 0.3$, yielding $\bar{H}^{\alpha}=\bar{H}^{\beta}\approx 5.34$.
\section{Conclusions and open issues}
\label{sect:extensions}
This paper presented a method for calculating the AoI in quasi-reversible, overtake-free networks of queues. The presented analysis simplified and extended the original analysis in \cite{kaul2012real} for FCFS single-server queues.

Results for $n$ identical M/M/1 queues in tandem showed that, although the minimum AoI value (achieved for some intermediate load value, which decreases with $n$) does not increase significantly with the number of queues, the effect of the added queues becomes apparent for high load values, and the age can then increase tremendously. Results for a simple network with two classes of update packets, entering through different queues in the network but sharing the output queue, revealed that changing the arrival rate of one class always effects on the AoI of that class, but the other class is significantly affected only when the output queue is close to saturation.

The use of the method for calculating the AoI requires that queues in the network path are Markovian and quasi-reversible, the path is overtake-free, and that the sojourn and interarrival time distributions at each queue are known. Quasi-reversibility for Markovian queues is equivalent to the Poisson-in-Poisson-out property and holds more generally for M/M/c queues with exponential constant-rate arrivals, as well as for processor-sharing queues in BCMP networks. 
However, these queues are not overtake-free; it remains an open issue to derive the sojourn time distribution of packets that are not overtaken in such queues (which are the ones that contribute fresh information and hence matter for the calculation of the AoI), as well as to find the probability that a packet is not overtaken by any of the subsequent packets. This would enable to extend the analysis presented in the paper. 
\bibliographystyle{plain}
\bibliography{aoi}

\begin{thebibliography}{10}

\bibitem{costa2014age}
Maice Costa, Marian Codreanu, and Anthony Ephremides.
\newblock Age of information with packet management.
\newblock In {\em 2014 IEEE International Symposium on Information Theory},
  pages 1583--1587. IEEE, 2014.

\bibitem{huang2015optimizing}
Longbo Huang and Eytan Modiano.
\newblock Optimizing age-of-information in a multi-class queueing system.
\newblock In {\em 2015 IEEE International Symposium on Information Theory
  (ISIT)}, pages 1681--1685. IEEE, 2015.

\bibitem{kam2018age}
Clement Kam, Sastry Kompella, Gam~D Nguyen, Jeffrey~E Wieselthier, and Anthony
  Ephremides.
\newblock On the age of information with packet deadlines.
\newblock {\em IEEE Transactions on Information Theory}, 64(9):6419--6428,
  2018.

\bibitem{kaul2012real}
Sanjit Kaul, Roy Yates, and Marco Gruteser.
\newblock Real-time status: How often should one update?
\newblock In {\em 2012 Proceedings IEEE INFOCOM}, pages 2731--2735. IEEE, 2012.

\bibitem{kaul2020timely}
Sanjit~K Kaul and Roy~D Yates.
\newblock Timely updates by multiple sources: The {M}/{M}/1 queue revisited.
\newblock In {\em 2020 54th Annual Conference on Information Sciences and
  Systems (CISS)}, pages 1--6. IEEE, 2020.

\bibitem{kaul2012status}
Sanjit~K Kaul, Roy~D Yates, and Marco Gruteser.
\newblock Status updates through queues.
\newblock In {\em 2012 46th Annual Conference on Information Sciences and
  Systems (CISS)}, pages 1--6. IEEE, 2012.

\bibitem{koukou20}
I.~{Koukoutsidis}.
\newblock Age of information in an overtake- free network of quasi - reversible
  queues.
\newblock In {\em 2020 28th International Symposium on Modeling, Analysis, and
  Simulation of Computer and Telecommunication Systems (MASCOTS)}, pages 1--6,
  2020.

\bibitem{melamed1982sojourn}
Benjamin Melamed.
\newblock Sojourn times in queueing networks.
\newblock {\em Mathematics of Operations Research}, 7(2):223--244, 1982.

\bibitem{nelson1993mathematics}
Randolph~D Nelson.
\newblock The mathematics of product form queuing networks.
\newblock {\em ACM Computing Surveys (CSUR)}, 25(3):339--369, 1993.

\bibitem{walrand1988introduction}
Jean Walrand.
\newblock {\em An introduction to queueing networks}.
\newblock Prentice Hall, 1988.

\bibitem{wu2017optimal}
Xianwen Wu, Jing Yang, and Jingxian Wu.
\newblock Optimal status update for age of information minimization with an
  energy harvesting source.
\newblock {\em IEEE Transactions on Green Communications and Networking},
  2(1):193--204, 2017.

\bibitem{yates2018age}
Roy~D Yates.
\newblock Age of information in a network of preemptive servers.
\newblock In {\em IEEE INFOCOM 2018-IEEE Conference on Computer Communications
  Workshops (INFOCOM WKSHPS)}, pages 118--123. IEEE, 2018.

\bibitem{yates2018networks}
Roy~D Yates.
\newblock The age of information in networks: Moments, distributions, and
  sampling.
\newblock {\em arXiv preprint arXiv:1806.03487}, 2018.

\bibitem{yates2012real}
Roy~D Yates and Sanjit Kaul.
\newblock Real-time status updating: Multiple sources.
\newblock In {\em 2012 IEEE International Symposium on Information Theory
  Proceedings}, pages 2666--2670. IEEE, 2012.

\bibitem{yates2018SHS}
Roy~D Yates and Sanjit~K Kaul.
\newblock The age of information: Real-time status updating by multiple
  sources.
\newblock {\em IEEE Transactions on Information Theory}, 65(3):1807--1827,
  2018.

\bibitem{yates2020age}
Roy~D Yates, Yin Sun, D~Richard Brown~III, Sanjit~K Kaul, Eytan Modiano, and
  Sennur Ulukus.
\newblock Age of information: An introduction and survey.
\newblock {\em arXiv preprint arXiv:2007.08564}, 2020.

\end{thebibliography}
\end{document}